\documentclass[twocolumn,showpacs,amsmath,amssymb,prl,superscriptaddress,floatfix,groupedaddress]{revtex4-1}
\pdfoutput=1

\usepackage{graphicx}
\usepackage{amssymb,amsfonts,amsmath}
\usepackage{hyperref}

\begin{document}

\title{Using the $s$-Ensemble to Probe Glasses Formed by Cooling and Aging}

\author{Aaron S. Keys$^{1,2}$, David Chandler$^1$ and Juan P. Garrahan$^3$} 
\email[Corresponding author, ]{Juan.Garrahan@nottingham.ac.uk}
\affiliation{$^{1}$Department of Chemistry, University of California, Berkeley CA, 94720}
\affiliation{$^{2}$Lawrence Berkeley National Laboratory, Berkeley CA, 94720}
\affiliation{$^{3}$School of Physics and Astronomy, University of Nottingham, Nottingham, NG7 2RD, United Kingdom}

\date{\today}

\begin{abstract}
From length scale distributions characterizing frozen amorphous domains, we relate the $s$-ensemble method with standard cooling and aging protocols for forming glass.  We show that in a general class of models, where space-time scaling is in harmony with that of experiment, the domain size distributions obtained with the $s$-ensemble are identical to those obtained through cooling or aging, but the computational effort for applying the $s$-ensemble is generally many orders of magnitude smaller than that of straightforward numerical simulation of cooling or aging. 
\end{abstract}

\maketitle

Through biasing statistics of trajectory space, the so-called ``$s$-ensemble'' method, non-equilibrium phase transitions emerge between ergodic liquid-like states and dynamically inactive glass-like states.  This class of transitions are found in idealized lattice models~\cite{merolle2005space, garrahan2007dynamical} and in simulations of atomistic models~\cite{hedges2009dynamic, speck2012constrained, speck2012first}.  In the latter case, it affords a systematic computational means of preparing exceptionally stable glass-states states~\cite{jack2011preparation}. This Letter draws the conclusion that the $s$-ensemble transition coincides with the physical glass transition \cite{reviews}, and the ensemble of its inactive states are those of natural structural glass.  Specifically, we derive correspondence between spatial correlations in the $s$-ensemble glass with those in the glass produced with finite-rate cooling or aging.  The correspondence provides a basis for an extraordinarily  efficient route for preparing structural glass with molecular simulations.

\begin{figure*}
\begin{center}
\includegraphics[width=0.9\textwidth]{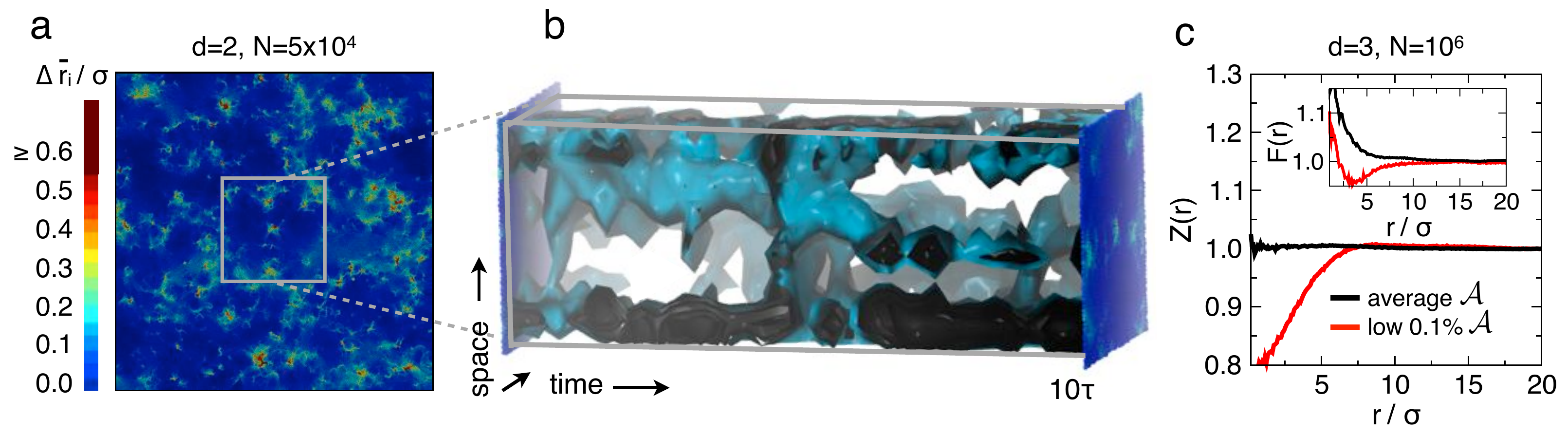}
\caption{\label{fig:fig1} 
Soft spots and space-time bubbles. (a) Burnished excitations for the $d=2$ supercooled WCA liquid mixture~\cite{keys2011excitations}, with $\Delta \bar{r}_i = [|\bar{\mathbf{r}}_i(\Delta t) - \bar{\mathbf{r}}_i(0)|]_\mathrm{iso}$.   (b) Excitation lines for a subsection of the system shown in (a).  (c) The functions $Z(r)$ and $F(r)$ demonstrating correlation holes for inactive subsystems of a supercooled $d=3$ WCA liquid mixture~\cite{hedges2007decoupling}.  
}
\end{center}
\end{figure*}

\bigskip
\noindent
\textbf{Space-time structure of glass-forming liquids.} To begin, it is helpful to consider
Figs.~\ref{fig:fig1}a and \ref{fig:fig1}b, which render trajectories of a two-dimensional $5\times10^4$-particle system in a fashion that extends the approach of Ref.~\cite{keys2011excitations}.  The system is a liquid mixture at a temperature that is 80\% below that of the onset temperature, $T_\mathrm{o}$~\cite{keys2011excitations,garrahan2003coarse, elmatad2009corresponding,elmatad2010corresponding,Note1}, and the trajectory runs for an observation time $t_\mathrm{obs} \approx 10\,\tau$.  Here, $\tau$ stands for the equilibrium structural relaxation time.  It is about $10^5$ integration steps at this particular temperature, and $t_\mathrm{obs}$, being 10 times longer, provides ample opportunity to observe the nature of dynamic heterogeneity in the system.

Most motions in glass-forming liquids are irrelevant vibrations, and the amplitudes of most of those vibrations are similar in size to typical enduring displacements~\cite{keys2011excitations}.  Irrelevant vibrations can be filtered out by focusing on inherent structures~\cite{stillinger1982hidden}.  The set of particle positions at time $t$, $\{\mathbf{r}_i(t)\}$, evolves by molecular dynamics; the inherent structure, $\{\bar{\mathbf{r}}_i(t)\}$, is the position of the potential-energy minimum closest to $\{\mathbf{r}_i(t)\}$.  The renderings in Fig.~\ref{fig:fig1} refer to $[a(\mathbf{r},t)]_\mathrm{iso}$, where
\begin{equation}
\label{eq:activityfield}
a(\mathbf{r},t) = \sum_{i=1}^N |\bar{\mathbf{r}}_i(t+\Delta t) - \bar{\mathbf{r}}_i( t)|\,\delta(\mathbf{r} - \bar{\mathbf{r}}_i(t))\,.
\end{equation} 
Here, we take $\Delta t$ to be $10^3$ integration steps, which is roughly the average time to complete an enduring displacement of one atomic diameter~\cite{keys2011excitations}, and $[\cdots]_\mathrm{iso}$ indicates iso-configurational averaging, which averages many trajectories of length $\Delta t$, all starting from the same configuration~\cite{widmer2004reproducible}.  Particles colored red in Fig.~\ref{fig:fig1}a are those for which the iso-configuration averaged $\bar{\mathbf{r}}_i(\Delta t)$ is more than $0.6\, \sigma$ from $\bar{\mathbf{r}}_i(0)$, where $\sigma$ is a particle diameter.  Smaller enduring displacements are colored by interpolating between red and blue, as noted in the color scale.  The figure thus shows that enduring displacements occurring over a short period of time, $\Delta t$, take place in sparse localized regions of space.  These regions are the excitations \cite{keys2011excitations} in an otherwise rigid material. 

Similar pictures illustrating arrangements of enduring displacements are found without iso-configuration averaging.  See Fig. 1 of Ref.~\cite{keys2011excitations} and also Media 1 of Ref.~\cite{keys2011excitations}.  Iso-configurational averaging serves to burnish those pictures \cite{Note2}.
Without that averaging, localized excitations are already evident but more irregular.  Importantly, the excitations, often referred to as soft spots~\cite{manning2011vibrational}, change little in size as the liquid is cooled, and further, there are no inter-excitation correlations at equal times~\cite{keys2011excitations}.  

Correlations develop over time through dynamics, which is illustrated with Fig.~\ref{fig:fig1}b.  That picture shows constant-value surfaces of $[a(\mathbf{r},t)]_\mathrm{iso}$. The surfaces form connected tubes or lines in space time -- excitation lines \cite{Garrahan2002} -- indicating that excitations facilitate birth (and death) of adjacent excitations.  Larger amplitude fluctuations of the surface occur less frequently than smaller amplitude fluctuations, indicating that the facilitated dynamics is hierarchical~\cite{palmer1984models}.  Lowering temperature reduces the number of excitations or soft spots, which reduces the probability that soft spots can connect, which reduces the rate at which the system can relax.

This behavior is found consistently in glass-forming liquids for all temperatures below the onset, i.e., $T<T_\mathrm{o}$~\cite{keys2011excitations}.  Throughout this regime, it is characterized by simple equations for space-time scaling and for the equilibrium distribution of distances between neighboring soft spots, $P_\mathrm{eq}(\ell)$:
\begin{equation}
\label{eq:spacetime}
\ell/\sigma = (\tau_\ell / \tau_\mathrm{o})^{1/\tilde{\beta}\gamma}\,.
\end{equation}
and
\begin{equation}
\label{eq:Peq}
 P_\mathrm{eq}(\ell) = \ell_{\mathrm{eq}}^{-1}\exp(-\ell/ \ell_\mathrm{eq})\,,  \quad    \ell_\mathrm{eq} = \sigma \exp(\tilde{\beta}/d_\mathrm{f})\,,
\end{equation}
Here, $\tilde{\beta} = J_\sigma /T - J_\sigma /T_\mathrm{o}$, where $J_\sigma$ is the energy of an excitation with enduring displacements of the characteristic structural length, $\sigma$, $1/\tau_\ell$ is the relaxation rate on length scale $\ell$, $d_\mathrm{f}$ is the fractal dimensionality of dynamic heterogeneity, and $\gamma$ is the proportionality constant for logarithmic growth of excitation energy with respect to length scale
\cite{Note3}.
The parabolic law~\cite{garrahan2003coarse, elmatad2009corresponding, keys2011excitations}, $\tau = \tau_\mathrm{o} \exp(\tilde{\beta}^2 \gamma / d_\mathrm{f})$, follows from space-time scaling, Eq.~(\ref{eq:spacetime}), evaluated at $\ell =\ell_\mathrm{eq}$.

Contributions to $P_\mathrm{eq}(\ell)$ with short inter-excitation lengths, $\ell < \ell_\mathrm{eq}$,  come from regions with excitation lines that connect and reorganize.  Contributions with $\ell \gg \ell_\mathrm{eq}$ come from regions of rigidity -- the empty regions of Fig.~\ref{fig:fig1}b, so-called ``bubbles'' in space-time \cite{Garrahan2002}.  When the liquid transforms into glass, the temporal extents of those bubbles grow to very long times, and excitation lines rarely or never touch, yielding a striped structure of trajectory space~\cite{keys2013calorimetric}.  In that case, excitations are no longer uncorrelated, and a non-equilibrium correlation length, $\ell_\mathrm{ne}$, gives the average or most probable separation of excitation lines.

Figure~\ref{fig:fig1}c shows that the equilibrium glass-forming liquid already contains the seeds of this non-equilibrium correlation length.
Specifically, for a $10^6$-particle WCA liquid mixture~\cite{hedges2007decoupling} in $d=3$ at $T=0.7\,T_\mathrm{o}$, Fig.~\ref{fig:fig1}c contrasts the equilibrium concentration of excitations with that surrounding a dynamically inactive sub-volume.  The trajectory length is $t_\mathrm{obs} = 50 \tau \approx 10^3 \Delta t$.
The net dynamical activity in a sub-volume $\Delta v$ is $\int_0^{t_\mathrm{obs}}dt \int_{\Delta v} d\mathbf{r}\, a(\mathbf{r},t)$.  For Fig.~\ref{fig:fig1}c, we have partitioned the total volume $V$ into cubes, each of size $\Delta v = 125 V/N$, and computed $Z(r) = \langle a(\mathbf{r}) \rangle_{\Delta v} / a$.  Here, $a$ is the equilibrium average of $a(\mathbf{r},t)$, and $\langle \cdots \rangle_{\Delta v}$ is that average conditioned on a low activity in the sub-volume at the origin.  Similarly, we have computed the radial distribution of mobility, $F(r) = \langle a(\mathbf{r}) \rangle_0 / a g(r)$, where $g(r) N/V$ is the mean particle density at $\mathbf{r}$ given a particle is at the origin and $\langle \cdots \rangle_0$ is the equilibrium average given a particle at the origin has just then completed an enduring displacement of at least $0.3\, \sigma$.  The red lines of Fig.~\ref{fig:fig1}c refer to the 0.1\% least active sub-volumes.  For $F(r)$, that means the central displacing particle is within such a low-activity sub-volume. The equilibrium $Z(r)$ exhibits no structure, and the equilibrium $F(r)$ decays over the length scale of a single excitation.  In contrast, the atypical low-activity functions show significant anti-correlation between neighboring excitations. 

\bigskip
\noindent
\textbf{Preparing glassy states.}  The statistical weight for these low-activity regions are enhanced by shifting to a non-equilibrium $s$-ensemble distribution, $P_s[x(t)] \propto P_0[x(t)] \exp(-s {\mathcal{A}}[x(t)])$ \cite{garrahan2007dynamical,hedges2009dynamic,speck2012constrained,Lecomte2007}. Here, $P_0[x(t)]$ is the equilibrium distribution functional for trajectories $x(t)$ of length $t_\mathrm{obs}$, and ${\mathcal{A}}[x(t)]$ is the net dynamical activity 
\cite{Note4},
\begin{equation}
{\mathcal{A}}[x(t)] = \int_V d\mathrm{r} \int_0^{t_\mathrm{obs}} dt\,\, a(\mathrm{r},t)\,.
\label{activity}
\end{equation}
For an ergodic equilibrium system, ${\mathcal{A}} = t_\mathrm{obs} V a$.  Deviations from this equilibrium value are measures of non-ergodic non-equilibrium behavior.  Time integrals of other quantities, not just the activity as in (4), can also serve as suitable order parameters to distinguish ergodic and non-ergodic behavior.  Time integration is the crucial feature.  Fluctuations are then intimately related to the behavior of time correlators.

Remarkably, for systems at $T<T_\mathrm{o}$, the marginal equilibrium distribution for ${\mathcal{A}}$ exhibits fat tails at low activity \cite{merolle2005space}, so that the non-equilibrium mean, $\langle \mathcal{A} \rangle_s$, changes abruptly around a transition value of $s$.  For $s<s^*$, the material is a normal melt, and for $s> s^*$, the material is an inactive amorphous phase -- a glass.  The abrupt change tends to a discontinuity as $Nt_\mathrm{obs} \rightarrow \infty$.  The glass transition in the $s$-ensemble is thus a first-order transition \cite{garrahan2007dynamical,hedges2009dynamic,speck2012constrained}.

\begin{figure*}
\begin{center}
\includegraphics[width=0.8\textwidth]{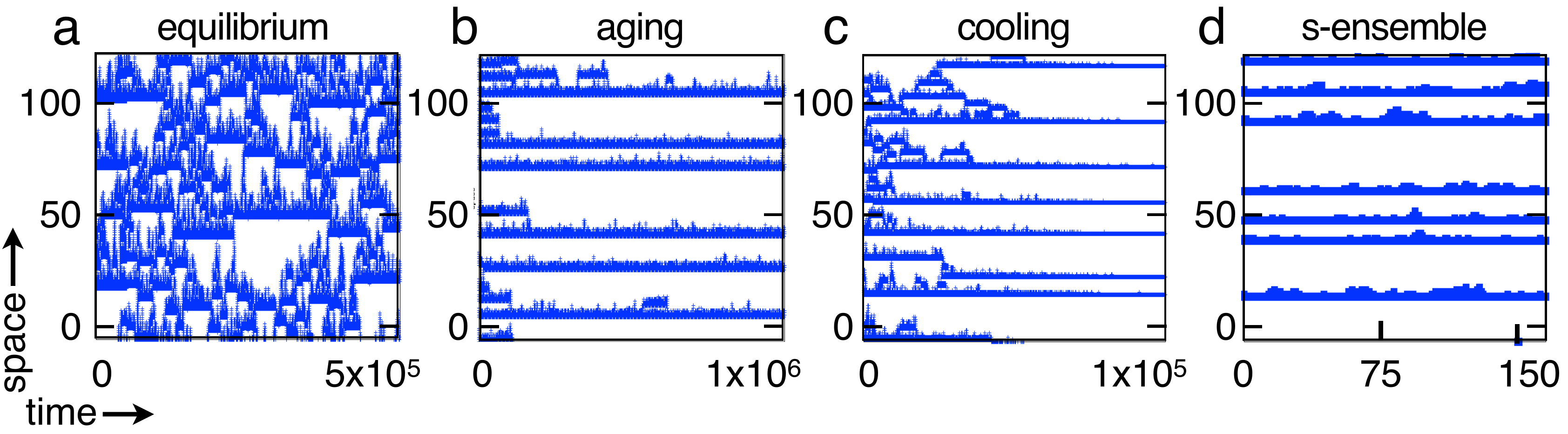}
\caption{\label{fig:fig2} 
Trajectories of excitations in the East model.  
(a) Equilibrium dynamics for $T=0.45$ over a time scale spanning about 50 structural relaxation times at that temperature.  (b) Aging dynamics after a quench from $T=1$ to $T=0.25$.  (c) Cooling at a rate $\nu=10^{-5}$. (d) Trajectory from the $s$-ensemble at $T=0.72$ and $s>s^* \approx 10^{-2}$, trajectories running for about 1/2 a structural relaxation time at that temperature. }
\end{center}
\end{figure*}

Even more remarkably, the transition can be obtained with $t_\mathrm{obs}$ much shorter than time scales required to produce glass from standard cooling protocols.  By cooling at a rate $\nu$, a glass transition occurs at the temperature $T_\mathrm{g}$, where $\nu^{-1} \approx |d \tau / dT|_{T= T_\mathrm{g}}$.  The time scale for that process is $\tau_\mathrm{g} = \tau(T_\mathrm{g})$.  The transition freezes excitations separated by the non-equilibrium length, $\ell_\mathrm{ne} = \ell_\mathrm{eq}(T_\mathrm{g})$.  From Eq.~(\ref{eq:spacetime}), $ \ell_\mathrm{ne}/\sigma = (\tau_\mathrm{g}/\tau_\mathrm{o})^{1/\tilde{\beta}_\mathrm{g}\gamma}$.
This length must be large if the glass persists for long times.  Thus, in view of Eq.~(\ref{eq:Peq}), $\tilde{\beta}_\mathrm{g} > 1$.  Typically, $\tau_\mathrm{g} > 10^{10} \tau_\mathrm{o}$ and $\ell_\mathrm{ne} \gtrsim 10 \sigma$.  

On the other hand, with the $s$-ensemble, the same large non-equilibrium length can be obtained with any positive value of $\tilde{\beta}$.  In that case, from Eq.~(\ref{eq:spacetime}), $\ell_\mathrm{ne}/ \sigma =  (t_\mathrm{obs}/\tau_\mathrm{o})^{1/\tilde{\beta} \gamma}$.  As such,
\begin{equation}
\label{eq:2times}
t_\mathrm{obs}/\tau_\mathrm{o} = (\tau_\mathrm{g} / \tau_\mathrm{o})^{\tilde{\beta} / \tilde{\beta}_\mathrm{g}}\,.
\end{equation}
The ratio $\tilde{\beta} / \tilde{\beta}_\mathrm{g}$ can be much smaller than 1.  In practice,  $\tilde{\beta} / \tilde{\beta}_\mathrm{g} \approx 1/10$.  Thus, the simulation time required to prepare a glassy state in the $s$-ensemble, $t_\mathrm{obs}$, is many orders of magnitude shorter than the time to prepare glass by straightforward cooling, $\tau_\mathrm{g}$.

\bigskip
\noindent
\textbf{Illustration with the East model.}  Equation~\ref{eq:2times} follows from well-tested scaling relationships, and there is some empirical evidence that glasses produced with the $s$-ensemble do indeed coincide with natural structural glass~\cite{jack2011preparation, limmer2013theory}.  Nevertheless, this relationship is not yet tested explicitly.  Here, we do so for the East model~\cite{eastmodel}, the simplest of models consistent with phenomenology of structural glasses and glass formers~\cite{elmatad2009corresponding,elmatad2012manifestations,keys2013calorimetric}.

\begin{figure}[b]
%\begin{center}
\includegraphics[width=0.95\columnwidth]{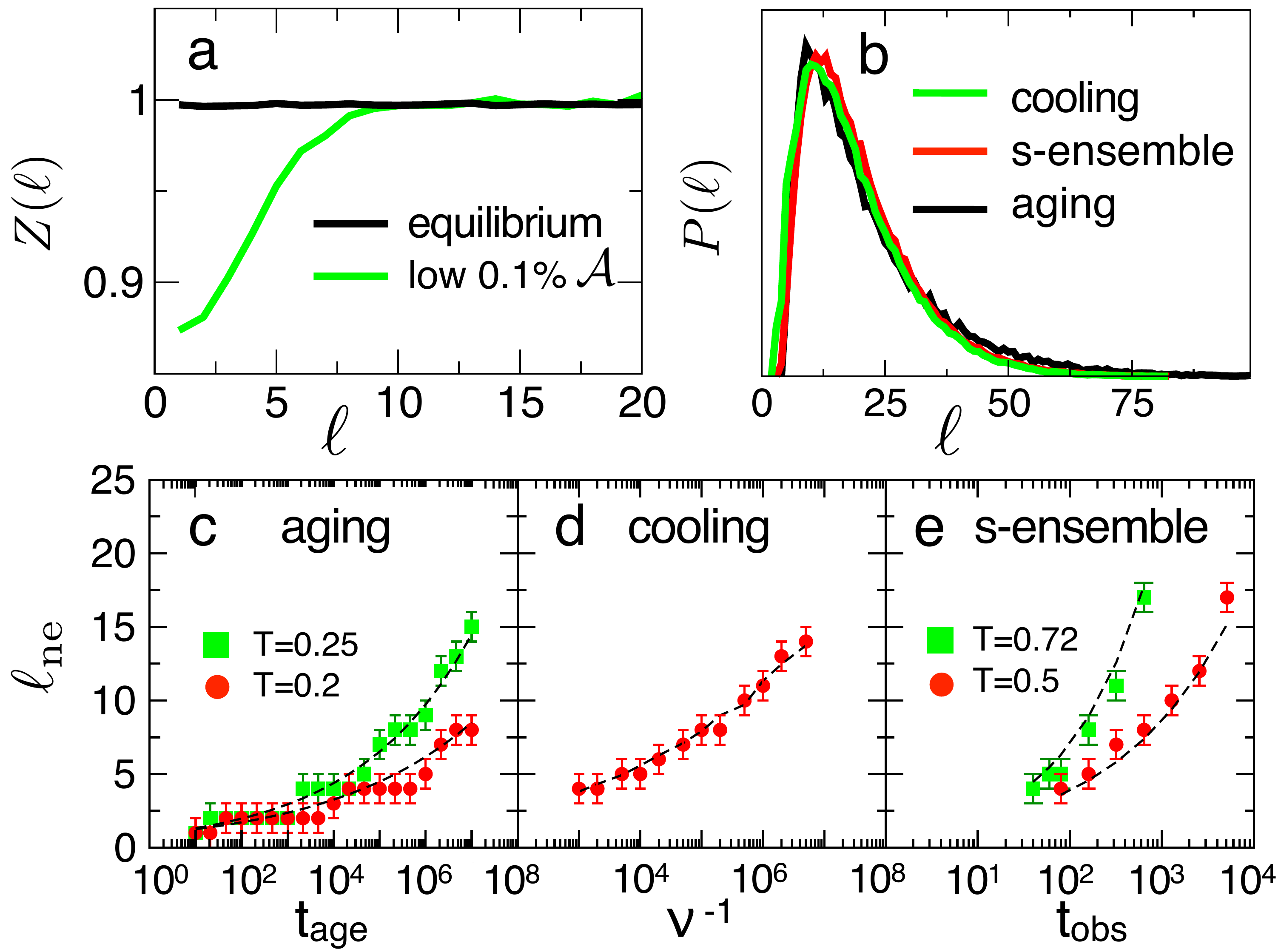}
\caption{\label{fig:fig3} Distributions of East-model glasses. (a) The relative concentration of excitations a distance $\ell$ from one of the least active sub-regions in the equilibrium model at temperature $T=0.5$  
%[LABEL ``cooling'' SEEMS INCORRECT.  ``least active''???]
(b) Distributions of domain lengths $P(\ell)$ for systems aged, cooled and driven with s, all designed to yield $\ell_\mathrm{ne} \approx 10$.  Here, $\ell$ represents the distance between frozen excitations (called ``super-spins''\cite{sollich2003glassy}).  Aging was done at $T=0.25$ for $t_\mathrm{age} = 2.5\times 10^{5}$ after quenching from $T=0.4$.  Cooling to $T=0$ was done with a cooling rate of $\nu = 10^{-6}$.  s-enemble trajectories were carried out at $T=0.72$ for a time duration of $t_\mathrm{obs}=320$.
(c-e) Growth of $\ell_\mathrm{ne}$ as a function of relevant time variables.  Dashed lines in (c) refer to $\ell_\mathrm{ne} =( t_\mathrm{age}/\tau_\mathrm{o})^{1/\tilde{\beta}\gamma}$; dashed line in (d) refers to $\ell_\mathrm{eq}(T_\mathrm{g})$; dashed lines in (e) refer to $\ell_\mathrm{ne}= ( t_\mathrm{obs}/\tau_\mathrm{o})^{1/\tilde{\beta}\gamma}$. 
}
%\end{center}
\end{figure}

In brief, the East model consists of a $d=1$ lattice with $N$ sites, each with variables $n_i = 0,1$.  The equilibrium concentration of excitations is $\langle n_i \rangle =c$.  At low temperatures, $c \sim \exp(-1/T)$.  (We take 1 as the energy scale and length scale for the model.)  Sites with $n_{i} = 1$ can facilitate a spin flip at the adjacent site $n_{i+1}$.  The corresponding transition rates are given for a site $i$ by $k_{i, 0 \rightarrow 1} =  n_{i-1} c/(1-c)$ and $k_{i, 1 \rightarrow 0} = n_{i-1}$.  The dynamics of this model is hierarchical~\cite{eastmodel, sollich2003glassy}.  Its structural relaxation slows by twelve orders of magnitude as $T$ decreases from 1 to 0.2~\cite{ashton2005fast}, it obeys space-time scaling of Eq.~(\ref{eq:spacetime}) and the parabolic law with $\gamma \approx 1/2 \ln 2$~\cite{ashton2005fast, aldous2002asymmetric, chleboun2013time}.  (A different value of $\gamma$ applies for aging regimes 
\cite{Note5}.)
An equilibrium trajectory of the model is shown in Fig.~\ref{fig:fig2}a.  Three protocols for preparing non-equilibrium glass states are illustrated in Figs.~\ref{fig:fig2}b, \ref{fig:fig2}c and \ref{fig:fig2}d.

In the first, aging, the model is initially equilibrated at $T=1$ and then instantly quenched to $T=0.25$, after which it runs at that the low temperature for times $t_\mathrm{age}$, where $\tau(1) \ll t_\mathrm{age} \ll \tau(0.25) \approx 3\times 10^{9}$.  During the time $t_\mathrm{age}$, the system can relax domains that are smaller than a characteristic length~\cite{sollich2003glassy, keys2013calorimetric} $\ell_\mathrm{ne} =  \left(t_\mathrm{age} / \tau_\mathrm{o} \right)^{1 / \tilde{\beta} \gamma}$, with $\tilde{\beta} \approx 3$.  We use $t_\mathrm{age} \approx 10^6$ so as to produce an average non-equilibrium spacing between excitations of about 10. 

In the second, cooling, the model is equilibrated at a temperature $T=1$ and then cooled to zero temperature at a rate of $\nu = 10^{-5}$.  A glass transition occurs at the stage where $1/\nu \approx |d\tau/dT|$, which gives $T_g \approx 0.48$ and thus $\ell_\mathrm{ne} \approx 10$ and $\tau_\mathrm{g} \approx 10^6$.  In other words, excitations in the glass are frozen in with a typical spacing of about 10, and the time scale to create the material is about $10^6$.  

The third case, the $s$-ensemble protocol, produces a similar glass in a much shorter time.  A similar inter-excitation distance is targeted with trajectories run at $T=0.72$ for which $\tau \approx 300$ and $\tilde{\beta} \approx 0.3$.  The glass transition from the cooling protocol occurs at $\tilde{\beta}_\mathrm{g} \approx 1/(1/2) -1 = 1$.  Accordingly, from Eq.~(\ref{eq:2times}), the $s$-ensemble transition for trajectories at $T=0.72$ produces the glass with $\ell_\mathrm{ne}$ when $t_\mathrm{obs} \approx 100$.  To apply the $s$-ensemble, we use the total number of enduring kinks as a measure of dynamical activity.  An enduring kink at site $i$ is a change in $n_i$ that persists for at least a mean exchange time~\cite{elmatad2012manifestations}.  At $T=0.72$ and $t_\mathrm{obs} \approx 100$, for system size $N$ chosen, the $s$-ensemble glass transition occurs at $s^* \approx 10^{-2} = O(1/N)$ \cite{Bodineau2012} (see Supplemental Material).

Figure \ref{fig:fig3} compares the non-equilibrium correlation lengths and distribution functions for the three different preparation protocols.  It also shows $Z(\ell)$, which is the relative concentration of enduring kinks a distance $\ell$ from the 0.1\% least-active domains of the equilibrium East model.  It exhibits a correlation hole in a fashion similar to the analogous $Z(r)$ in the WCA mixture, Fig.~\ref{fig:fig1}.  The East model thus illustrates how preparation of glass, which necessarily requires long physical times, can be accomplished in simulation in much shorter times through application of the $s$-ensemble.  Equation (\ref{eq:2times}) provides the key for understanding prior successes in preparing glassy states through applications of the $s$-ensemble in atomistic models.  However, if the simulation box size, $L$, is smaller than target non-equilibrium length, the $s$-ensemble method prepares a distribution of glassy states, all of which correspond to inactive domains in glasses with $\ell_{\rm ne} > L$.
  
Natural dynamics changes $\ell$ continuously, and at the point where the system falls out of equilibrium $\ell_{\rm ne} = \ell_{\rm eq}$.  The equilibrium length, $ \ell_{\rm eq}=\langle \ell \rangle$, is equivalent to $a$ or $c$.  Because the length scale changes continuously as a glass former falls out of equilibrium, the glass transition has the appearance of a second-order transition.  However, the time-integrated order parameters, the distribution of $\ell$, and the connection between $\ell$ and $c$, all change abruptly.  The change becomes singular in the limit of infinite time, manifesting the first-order non-equilibrium transition that underlies the glass transition.

\bigskip
\noindent
\textbf{Acknowledgements.}  We thank D.T. Limmer, R.L. Jack, P. Sollich, T. Speck and Y.S. Elmatad for helpful discussions.  Salaries were supported by the Director, Office of Science, Office of Basic Energy Sciences, and by the Division of Chemical Sciences, Geosciences, and Biosciences of the U.S. Department of Energy at LBNL, by the Laboratory Directed Research and Development Program at Lawrence Berkeley National Laboratory under Contract No. DE-AC02-05CH11231, and by Leverhulme Trust grant no.\ F/00114/BG.  NSF award CHE-1048789 provided computational resources.

%\bibliography{preprint}

%merlin.mbs apsrev4-1.bst 2010-07-25 4.21a (PWD, AO, DPC) hacked
%Control: key (0)
%Control: author (8) initials jnrlst
%Control: editor formatted (1) identically to author
%Control: production of article title (-1) disabled
%Control: page (0) single
%Control: year (1) truncated
%Control: production of eprint (0) enabled
%

\appendix

\widetext

\section{Supplemental Material}

\begin{figure*}
\begin{center}
\includegraphics[width=0.8\textwidth]{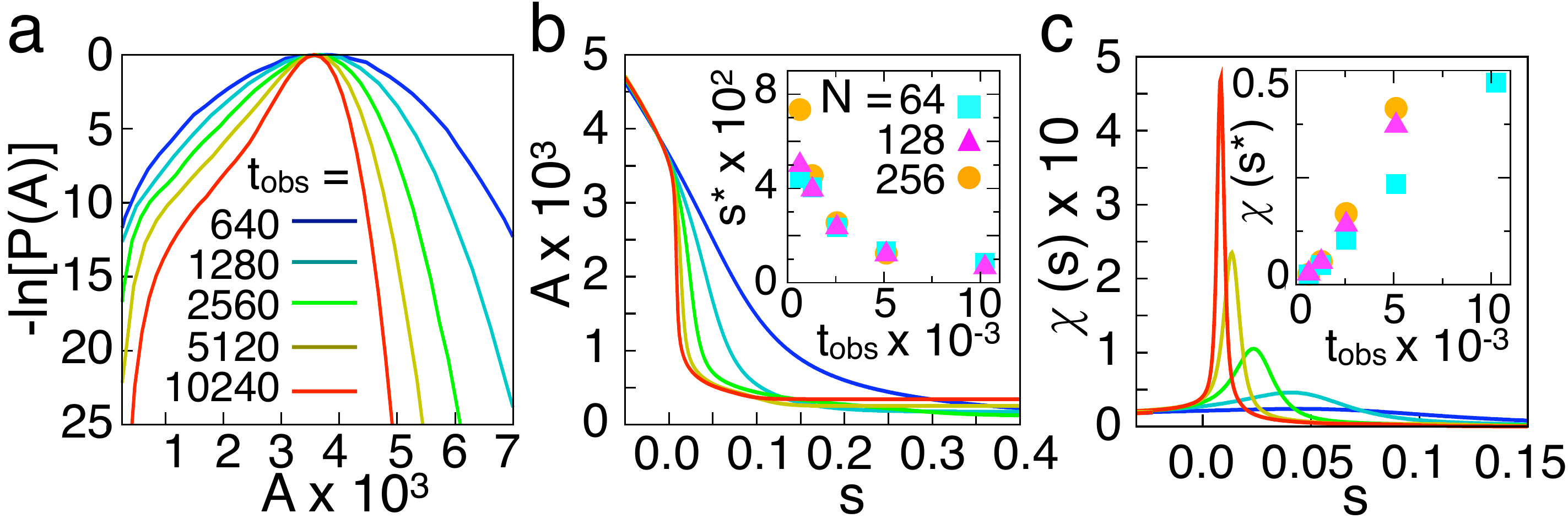}
\caption{\label{fig:fig4} 
First-order dynamical phase transition in the $s$-ensemble with activity measured in terms of enduring kinks.  (a) Probability of observing a trajectory with intensive activity $A$ as a function of $t_\mathrm{obs}$ for the $d=1$ East model at $T=0.72$ and $N=64$.  (b) $A$ as a function of $s$ for the same systems sampled in (a).  The value of $s^*$ is plotted as a function of $t_\mathrm{obs}$ in the inset for different system sizes $N$.  (c) Susceptibility $\chi(s)$ as a function of $s$.  The inset shows the peak of the susceptibility, $\chi(s^*)$ as a function of $t_\mathrm{obs}$ for different $N$.}
\end{center}
\end{figure*}

Aspects of the $s$-ensemble results are detailed in Fig.~\ref{fig:fig4}. The $s$-ensemble is sampled according to the methods outlined in Refs.~\cite{jack2006space, speck2012constrained}.  We use standard transition path sampling with both shooting and shifting moves to sample trajectory space.  The $s$-ensemble at each state point is sampled within 20 simulation windows, $w$, each with a different target value of activity, $\mathcal{A}_w$.  Trajectories are accepted or rejected according to a standard umbrella sampling criterion~\cite{torrie1977nonphysical}, with a harmonic biasing potential acting on the activity for each window, $W = \ k \left(\mathcal{A}[X] - \mathcal{A}_w \right)^2$, where $\mathcal{A}[X]$ is the total number of enduring kinks for the trajectory $X$.  At the state points considered, optimal sampling is obtained for $k \approx 10^5$.  Replica exchange between windows is implemented to facilitate the sampling of glassy states with low activity, which are inherently slowly-evolving.  Un-biased statistical averages are obtained using the multi-state Bennet acceptance ratio method~\cite{shirts2008statistically}. 

Whereas aging tends to eliminate short domains because larger domains are kinetically frozen, the $s$-ensemble eliminates short domains because of the statistical penalty imposed by the field $s$.  In both cases, domains that are shorter than $\ell_\mathrm{ne}$ relax on average while larger domains remain intact.  For aging, $\ell_\mathrm{ne} \sim t_\mathrm{age}^{1/\tilde{\beta} \gamma}$, where $\tilde{\beta}$ coincides with the temperature of the quench.  For the $s$-ensemble, $\ell_\mathrm{ne} \sim t_\mathrm{obs}^{1/\tilde{\beta} \gamma}$, where $\tilde{\beta}$ coincides with the temperature of the $s$-ensemble trajectories. 

The probability density of intensive activity, $A \equiv \mathcal{A}[X] / N t_\mathrm{obs}$,  is plotted in Fig.~\ref{fig:fig4}(a) as a function of $t_\mathrm{obs}$ for $s=0$.  For all $t_\mathrm{obs}$, the activity distribution is non-Gaussian and exhibits a fat tail for low values of activity.  This is the signature of a low-activity phase that can be accessed by driving the system with $s$.  Fig.~\ref{fig:fig4}(b) shows that, for $s$ exceeding a critical value $s^*$ (i.e., the value of $s$ that maximizes $\mathrm{d}A/\mathrm{d}s$), the system undergoes a phase transition into this inactive state.  The value of $s^*$ tends to zero as $t_\mathrm{obs} \rightarrow \infty$, as $\mathcal{A}[X]$ grows extensively with time.  The sharpness of the transition is quantified by a susceptibility $\chi(s) \equiv \mathrm{d}A / \mathrm{d} s = \left<A^2 \right> - \left< A \right>^2$, plotted in Fig.~\ref{fig:fig4}(c).  The length $\ell_\mathrm{ne}$ exceeds the system size $N$ in the limit $t_\mathrm{obs} \rightarrow \infty$, and the system forms a single domain of length $N$.  (The final spin cannot be eliminated due to the boundary conditions and facilitation rules.)  Activity fluctuations at $s=s^*$ therefore scale proportionally with $N t_\mathrm{obs}$, as illustrated in the inset of Fig.~\ref{fig:fig4}(c).  This scaling is the hallmark of a first-order dynamical phase transition at $s=s^*$.  If $\ell_\mathrm{ne}$ exceeds $N$, the system undergoes a first order transition to an ideal inactive phase; otherwise, the transition is smooth and the system falls into a striped phase.

\end{document}